\documentclass{article}

\def \lket {|}
\def \rket {\rangle}
\def \lbra {\langle}
\def \rbra {|}
\def \qed {\framebox(6,6)\\}
\def \M{{\cal M}}
\def \K{{\cal K}}
\def \H{{\cal H}}
\newcommand{\ket}[1]{\lket #1\rket}
\newcommand{\bra}[1]{\lbra #1\rbra}
\newtheorem{Definition}{Definition}
\newtheorem{Theorem}{Theorem}
\newtheorem{Lemma}{Lemma}

\newtheorem{Corollary}{Corollary}

\newcommand\comment[1]{}
\def\bbbc{{\mathchoice {\setbox0=\hbox{$\displaystyle\rm C$}\hbox{\hbox
to0pt{\kern0.4\wd0\vrule height0.9\ht0\hss}\box0}}
{\setbox0=\hbox{$\textstyle\rm C$}\hbox{\hbox
to0pt{\kern0.4\wd0\vrule height0.9\ht0\hss}\box0}}
{\setbox0=\hbox{$\scriptstyle\rm C$}\hbox{\hbox
to0pt{\kern0.4\wd0\vrule height0.9\ht0\hss}\box0}}
{\setbox0=\hbox{$\scriptscriptstyle\rm C$}\hbox{\hbox
to0pt{\kern0.4\wd0\vrule height0.9\ht0\hss}\box0}}}}

\begin{document}

\title{A New Protocol and Lower Bounds for Quantum Coin Flipping}

\author{Andris Ambainis\thanks{Supported
  by NSF Grant CCR-9987845 and the State of
  New Jersey. Part of this work done while at University of California,
  Berkeley (supported by Microsoft Research Graduate Fellowship and
	 NSF grant CCR-9800024) and IBM Almaden.}\\
       School of Mathematics\\
       Institute for Advanced Study\\
       Princeton, NJ 08540\\
       e-mail: {\tt ambainis@ias.edu}}

\date{}

\maketitle                                       

\begin{abstract}
We present a new protocol and two lower bounds for quantum coin flipping.
In our protocol, no dishonest party can achieve one outcome 
with probability more than 0.75. 
%(The previous best result was 0.91...\cite{ATVY}.) 
Then, we show that our protocol is optimal for a certain type of
quantum protocols.

For arbitrary quantum protocols, 
we show that if a protocol achieves a bias
of at most $\epsilon$, it must use at least 
$\Omega(\log \log \frac{1}{\epsilon})$ rounds of communication.
This implies that the parallel repetition fails for quantum coin
flipping.
(The bias of a protocol cannot be arbitrarily
decreased by running several copies of it in parallel.) 
\end{abstract}

\section{Introduction}

In many cryptographic protocols, there is a need for random bits
that are common to both parties. However, if one of parties is 
allowed to generate these random bits, this party may have a chance
to influence the outcome of the protocol by appropriately picking
the random bits. This problem can be solved by using a cryptographic
primitive called {\em coin flipping}.

\begin{Definition}
\label{SecDef}
A coin flipping protocol with $\epsilon$ bias is one where Alice and Bob
communicate and finally decide on a value $c\in\{0, 1\}$ such that
\begin{itemize}
\item
If both Alice and Bob are honest, then
$Prob(c=0)=Prob(c=1)=1/2$.
\item
If one of them is honest (follow the protocol), 
then, for any strategy of the dishonest player,
$Prob(c=0)\leq 1/2+\epsilon$, $Prob(c=1)\leq 1/2+\epsilon$.
\end{itemize}
\end{Definition}

Classically, coin flipping was introduced by Blum\cite{Blum}.
Classical coin flipping protocols are based on computational 
assumptions such as one-way functions.

However, classical one-way functions may not be hard against
quantum adversaries. (For example, factoring and discrete log are
not hard in the quantum case\cite{Shor}.) 
Finding a good candidate for
a one-way function secure against quantum adversaries is an 
important open problem.

On the other hand, the unique properties of
quantum mechanics allow the implementation of certain
cryptographic tasks without any computational assumptions.
(The security proof is based only on the
validity of quantum mechanics.) The most famous example is
the quantum key distribution\cite{BB,Sec1,Sec2,Sec3,Sec4}.
The question is: can we replace the computational assumptions
of the classical case by information-theoretic security
in the quantum case for the coin flipping?

For bit commitment (a related cryptographic primitive), this is
impossible\cite{LC97,LC,Mayers}\footnote{It is possible,
however, to have quantum protocols for bit commitment
under quantum complexity assumptions (existence of quantum
1-way functions). See Dumais et.al.\cite{Dumais} and 
Crepeau et.al.\cite{CLS}}.
The ideas of this impossibility proof can be used to
show that there is no quantum protocol for perfect quantum 
coin flipping (quantum coin flipping with bias 0) \cite{LC,MSC}.
However, this still leaves the possibility that there might be
quantum protocols with an arbitrarily small bias $\epsilon>0$.

\comment{Several protocols for quantum coin flipping have been proposed.
The first was the protocol by Goldenberg et.al.\cite{Goldenberg}
who used a weaker definition of security\footnote{Namely, 
\cite{Goldenberg} assumes that it is known in advance that 
Alice wants to bias the coin to 0 and Bob wants to bias it to 1.
Then, it is enough to give guarantees about $Pr[c=0]$ if
Bob is honest but Alice cheats and $Pr[c=1]$ if Alice is honest
but Bob cheats. In contrast, our definition \ref{SecDef} requires
that neither of players can bias the coin in any direction by more
than $\epsilon$.} and gave a protocol in which any dishonest 
player can achieve his desired outcome with probability at most 0.827...}

The first positive result was by
Aharonov et.al.\cite{ATVY} who gave a protocol for quantum coin flipping
%with the stronger security
%guarantee of Definition \ref{SecDef}.   
%In their protocol, 
in which a dishonest party cannot force a given 
outcome with probability more than 0.9143...
%A different protocol (with probability of successful cheating at most
%0.87...) was constructed by ....
%This result uses a simple protocol and gives provable
%guarantees about its security.

There has been some effort to construct more complicated protocols 
which would achieve arbitrarily small $\epsilon>0$.
At least two protocols have been proposed: by Mayers et.al.\cite{MSC} 
and by Zhang et.al.\cite{Zhang}. None of them had provable security 
guarantees but both were conjectured to achieve an arbitrarily small
$\epsilon>0$ for an appropriate choice of parameters.
Both of them were eventually broken: the protocol of \cite{MSC}
was broken by \cite{Gottesman,Leslau,Tokunaga}\footnote{The paper 
\cite{Tokunaga} claims to break any protocol for coin flipping
but this claim is incorrect. It does break a class of protocols
which includes the one of \cite{MSC}, though.}
and the protocol of \cite{Zhang} is insecure 
because of our Theorem \ref{Rounds}.

In this paper, we give a simple protocol in which a dishonest party
cannot achieve one outcome with probability more than 0.75.

Then, we show that our protocol is optimum in a certain class of protocols
that includes our protocol, the protocol of \cite{ATVY} and
other similar protocols.

Our third result (Theorem \ref{Rounds})
shows that, if there is a protocol with an arbitrarily 
small bias $\epsilon>0$, it must use a non-constant number of
rounds of communication
%then one needs to use more and more 
%rounds of communication 
({\em not just communicate many qubits in a
constant number of rounds}). 
Namely, a coin flipping algorithm
with a bias $\epsilon$ needs to have at least 
$\Omega(\log \log \frac{1}{\epsilon})$ rounds.
In particular, 
this means that the parallel repetition fails for quantum coin flipping.
(One cannot decrease the bias arbitrarily by repeating the protocol 
in parallel many times.)
%This also shows that the conjecture of \cite{Zhang} that
%their 3-round protocol achieves an arbitrarily small $\epsilon>0$ if
%sufficiently many qubits are transmitted in each round is incorrect.

{\bf Related work.}
We have recently learned that two 
of results in this paper (the 0.75 protocol and the matching 
lower bound for a class of protocols) have been independently
discovered by Spekkens and Rudolph \cite{RS0}.
Also, Kitaev \cite{Kitaev} has very recently shown that, in any 
protocol, at least one party can achieve one 
outcome with probability at least $1/\sqrt{2}=0.71...$. 
Thus, our 0.75 protocol is close to being optimal.

Curiously, Kitaev's lower bound does not apply to
a weaker version of coin flipping. In {\em weak coin flipping},
it is known in advance that 
Alice wants to bias the coin to 0 and Bob wants to bias it to 1.
Then, it is enough to give guarantees about $Pr[c=0]$ if
Bob is honest but Alice cheats and $Pr[c=1]$ if Alice is honest
but Bob cheats.
Protocols for weak coin flipping have 
been studied by Goldenberg \cite{Goldenberg}, 
Spekkens and Rudolph \cite{RS} and Ambainis\cite{Ambainis}.
The best protocol \cite{RS} achieves a maximum bias of
$1/\sqrt{2}$. We note best lower bound for weak coin
flipping is Theorem \ref{Rounds} of this paper.
% applies to weak coin flipping and, currently,
%it is the best lower bound known for this problem. 

The role of rounds in quantum communication has been studied
in a different context (quantum communication complexity
of pointer jumping) by Klauck et. al.~\cite{NTZ}.
There is a popular survey of quantum cryptography by Gottesman and 
Lo\cite{GL}.

\section{Preliminaries}

\subsection{Quantum states}

We briefly introduce the notions used in this paper.
For more detailed explanations and examples, see \cite{NC}.

\begin{description}
\item[Pure states:]
An $n$-dimensional pure quantum state is a vector $\ket{\psi}\in\bbbc^n$ of 
norm 1.
Let $\ket{0}$, $\ket{1}$, $\ldots$, $\ket{n-1}$ be an orthonormal basis for
$\bbbc^n$. Then, any pure state can be expressed 
as $\ket{\psi}=\sum_{i=0}^{n-1} a_i \ket{i}$ for some $a_0\in\bbbc$,
$a_1\in\bbbc$, $\ldots$, $a_{n-1}\in\bbbc$.
Since the norm of $\ket{\psi}$ is 1, $|a_i|^2=1$. 

The simplest special case is $n=2$. 
Then, the basis for $\bbbc^2$ consists of two vectors $\ket{0}$ and
$\ket{1}$ and any pure state is of form $a\ket{0}+b\ket{1}$,
$a\in\bbbc$, $b\in\bbbc$, $|a|^2+|b|^2=1$.
Such quantum system is called a {\em quantum bit (qubit)}.

We often look at $\ket{\psi}$ as a column vector consisting of 
coefficients $a_i$.
Then, we use $\bra{\psi}$ to denote the conjugate transpose of $\ket{\psi}$.
$\bra{\psi}$ is a row vector consisting of $a^*_i$ (complex conjugates of
$a_i$).
In this notation, $\lbra\psi|\phi\rket$ denotes the inner product of 
$\psi$ and $\phi$. (If $\ket{\psi}=\sum a_i \ket{i}$, 
$\ket{\phi}=\sum b_i \ket{i}$, then $\lbra \psi|\phi\rket=\sum a^*_i b_i$.)
$\ket{\psi}\bra{\phi}$ denotes the outer product of $\psi$ and $\phi$
(an $n\times n$ matrix with entries $a_i b^*_j$).

\item[Mixed states:]
A mixed state is a classical probability distribution 
$(p_i, \ket{\psi_i})$, $0\leq p_i\leq 1$, $\sum_i p_i=1$ 
over pure states $\ket{\psi_i}$.
The quantum system described by a mixed state
is in the pure state $\ket{\psi_i}$ with probability $p_i$.

A mixed state can be also described by its density matrix
$\rho=\sum_i p_i\ket{\psi_i}\bra{\psi_i}$.
It can be shown that any density matrix has trace 1.
({\em A trace} of a matrix is the sum of its diagonal entries.)

A quantum system can undergo two basic operations:
a unitary evolution and a measurement.

\item[Unitary evolution]:
A {\em unitary transformation} $U$
is a linear transformation on $\bbbc^{k}$ 
that preserves the $l_2$ norm (i.e., maps vectors of unit norm
to vectors of unit norm).

If, before applying $U$, the system was in a pure state $\ket{\psi}$,
then the state after the transformation is 
$U \ket{\psi}$. 

If, before $U$, the system was in a mixed state 
with a density matrix $\rho$, the state after the transformation
is the mixed state with the density matrix $U\rho U^{\dagger}$.

\item[Projective measurements]:
An observable is a decomposition of $\bbbc^k$ into
orthogonal subspaces $\H_1$, $\ldots$, $\H_l$: 
$\bbbc^n=\H_1\oplus \H_2\oplus \ldots \oplus \H_l$.
A measurement of a pure state $\ket{\psi}$ with respect
to this observable gives the result $i$ with probability 
$\|P_i \ket{\psi} \|^2$ where $P_i\ket{\psi}$ denotes the projection of 
$\ket{\psi}$ to the subspace $\H_i$. 
After the measurement, the state of the system becomes
$\frac{P_i\ket{\psi}}{\|P_i\ket{\psi}\|}$.

A more general class of measurements are POVM measurements 
(see \cite{NC}).
In most of this paper, it will be sufficient to consider 
projective measurements.
\end{description}

\subsection{Bipartite states}

\begin{description}
\item[Bipartite states:]
In the analysis of quantum coin flipping protocols, we 
often have a quantum state part of which is held by Alice and the other
part by Bob. For example, we can have the EPR state (the state of two qubits
$\frac{1}{\sqrt{2}}\ket{0}\ket{0}+\frac{1}{\sqrt{2}}\ket{1}\ket{1}$),
with the first qubit held by Alice and the second qubit held by Bob.
Such states are called {\em bipartite} states.

\item[Tracing out:]
If Alice measures her part, Bob's part becomes a mixed state.
For example, if Alice measures the first qubit of the EPR state
in the basis consisting of $\ket{0}$ and $\ket{1}$, Bob's state
becomes $\ket{0}$ with probability 1/2 and $\ket{1}$ with probability 1/2.
Let $\rho$ be the density matrix of the mixed state that Bob gets
if Alice measures her part of a bipartite state $\ket{\psi}$.
Then, we say that $\rho$ is obtained by {\em tracing out} the Alice's part
of $\ket{\psi}$.

There are many different ways how Alice can measure (trace out) her part.
However, they all give the same density matrix $\rho$ for Bob's
part.

\item[Purification:]
Let $\rho$ be a mixed state.
Then, any pure state $\ket{\psi}$ of a larger system that gives
$\rho$ if a part of the system is traced out is called a {\em purification}
of $\rho$.
\end{description}

\subsection{Distance measures between quantum states}

We use two measures of distance between quantum states
(represented by density matrices): trace distance and fidelity. 
For more information on these (and other) measures of distance
between density matrices, see \cite{Fuchs,NC}.

\begin{description}
\item[Trace distance:]
Let $p=(p_1, \ldots, p_k)$ and 
$q=(q_1, \ldots, q_k)$ be two classical probability distributions.
Then, the {\em variational distance} between $p$ and $q$
is 
\[ |p-q|=\sum_{i=1}^k |p_i-q_i|. \]
The variational distance characterizes how well one
can distinguish the distributions $p$ and $q$.

In the quantum case, the counterpart of a probability distribution
is a mixed state.
The counterpart of the variational distance is the trace
distance. It is defined as follows.

The trace norm of a matrix $A$ is the trace of $|A|$ where
$|A|=\sqrt{A^{\dagger}A}$ is the positive square root of $A^{\dagger}A$.
We denote the trace norm of $A$ by $\|A\|_t$.
The following lemma relates the trace norm of $\rho_1-\rho_2$
(which we also call trace distance between $\rho_1$ and $\rho_2$)
with the variational distance between distributions obtained
by measuring $\rho_1$ and $\rho_2$.

\begin{Lemma}
\cite{AKN}
\label{AKNLemma}
Let $p_{\rho_1}^M$, $p_{\rho_2}^M$ be the probability distributions 
generated by applying a measurement $\M$ to mixed states $\rho_1$ and $\rho_2$.
Then, for any (projective or POVM) measurement
$\M$, $|p_{\rho_1}^M-p_{\rho_2}^M|\leq \|\rho_1-\rho_2\|_t$
and there exists a measurement $\M$ that achieves the variational
distance $\|\rho_1-\rho_2\|_t$.
\end{Lemma}

We can always choose the measurement $\M$ that achieves the variational
distance $\|\rho_1-\rho_2\|_t$ so that $\M$ is a projective measurement and 
it has just two outcomes: 0 and 1. 

%In particular, if two mixed states have the same density matrix,
%then $\|\rho_1-\rho_2\|_t=0$. This means that applying
%any measurement to $\rho_1$ and $\rho_2$ gives the same probability 
%distribution

\item[Fidelity:]
Let $\ket{\psi_1}$ and $\ket{\psi_2}$ be two bipartite states.
Let $\rho_1$ and $\rho_2$ be the mixed states obtained from
$\ket{\psi_1}$ and $\ket{\psi_2}$ by tracing out (measuring)
Alice's part.

\begin{Lemma}
\label{TTheorem}
\cite{LC,Mayers}
If $\rho_1=\rho_2$, then Alice can transform $\ket{\psi_1}$
into $\ket{\psi_2}$ by a transformation on her part of the state.
\end{Lemma}

For example, consider the bipartite states 
\[ \ket{\psi_1} = \frac{1}{\sqrt{2}} (\ket{00}+\ket{01}) ,\]
\[ \ket{\psi_2} = \frac{1}{2} (\ket{00}+\ket{01}+\ket{10}-\ket{11}) ,\]
with the first qubit held by Alice and the second qubit held
by Bob. If Alice measures her qubit of $\ket{\psi_1}$,
Bob is left with $\ket{0}$ with a probability 1/2 and
$\ket{1}$ with a probability 1/2.
If Alice measures her qubit of $\ket{\psi_2}$,
Bob is left with $\frac{1}{\sqrt{2}}(\ket{0}+\ket{1})$ 
with a probability 1/2 and
$\frac{1}{\sqrt{2}}(\ket{0}-\ket{1})$ 
with a probability 1/2.
Both of those states have the same density matrix
\[ \left(
\begin{array}{ccc}
1/2 & 0 \\
0 & 1/2 
\end{array} \right) . \]
By lemma \ref{TTheorem}, this means that
Alice can transform $\ket{\psi_1}$ into $\ket{\psi_2}$.
Indeed, she can do that by applying the Hadamard transform $H$ to her 
qubit. 

A generalization of lemma \ref{TTheorem} is:
if the two density matrices $\rho_1$ and $\rho_2$ are close,
then Alice can transform $\ket{\psi_1}$ into a state $\ket{\psi'_1}$
that is close to $\ket{\psi_2}$.

%The fidelity measures how well Alice can transform a bipartite state
%$\ket{\psi_1}$ to a bipartite state $\ket{\psi_2}$ by applying only 
%transformations on her part.

In this case, the distance between the two density matrices is 
measured by the fidelity $F(\rho_1, \rho_2)$.
The fidelity is defined as 
\[ F(\rho_1, \rho_2)=\max_{\ket{\psi_1}, \ket{\psi_2}} 
|\lbra \psi_1 |\psi_2\rket|^2, \] 
over all choices of $\ket{\psi_1}$ and $\ket{\psi_2}$
that give density matrices $\rho_1$ and $\rho_2$ when a part of system
is traced out.

\begin{Lemma}
\cite{Jozsa}
\label{JLemma}
Let $\rho_1$, $\rho_2$ be two mixed states with support in a 
Hilbert space $\H$, $\K$ any Hilbert space of dimension at least
$\dim(\H)$, and $\ket{\phi_i}$ any purifications of $\rho_i$
in $\H\otimes\K$. Then, there is a local unitary transformation $U$
on $\K$ that maps $\ket{\phi_2}$ to 
$\ket{\phi'_2}=I\otimes U\ket{\phi_2}$ such that
\[  |\lbra \phi_1|\phi'_2 \rket |^2 = F(\rho_1, \rho_2) . \]
\end{Lemma}

\begin{Lemma}
\label{ULemma}
\cite{Uhlmann}
\[ F(\rho_1, \rho_2)=\left[Tr\left(\sqrt{\sqrt{\rho_1}\rho_2\sqrt{\rho_1}}
\right)\right]^2 .\]
\end{Lemma}

\item[Relation between trace distance and fidelity:]
The trace distance and the fidelity are closely related.
If $\rho_1$ and $\rho_2$ 
are hard to distinguish for Bob, then Alice can transform 
$\ket{\psi_1}$ into a state close to $\ket{\psi_2}$
and vica versa. 
Quantitatively, this relation is given by

\begin{Lemma}
\cite{Fuchs}
\label{FLemma}
For any two mixed states $\rho_1$ and $\rho_2$,
\[ 1-\sqrt{F(\rho_1, \rho_2)} \leq \frac{1}{2} \| \rho_1-\rho_2\|_t
\leq \sqrt{1-F(\rho_1, \rho_2)} .\]
\end{Lemma}

In particular, $F(\rho_1, \rho_2)=0$ if and only if 
$\|\rho_1-\rho_2\|_t=2$. 

\end{description}

%In particular, if there is a measurement for Bob which distinguishes
%$\ket{\psi_1}$ and $\ket{\psi_2}$ perfectly (gives the result 1 
%with probability 1 on $\rho_1$ and 2 with probability 1 on $\rho_2$),
%then $\|\rho_1-\rho_2\|_t=2$ and, by Lemma \ref{

\section{A protocol with bias 0.25}
\label{section2}

{\bf Protocol:}
Define 
\[ \ket{\phi_{b, x}}=\left\{ \begin{array}{cc}
\frac{1}{\sqrt{2}} \ket{0} + \frac{1}{\sqrt{2}} \ket{1} 
& \mbox{if $b=0$, $x=0$} \cr
\frac{1}{\sqrt{2}}\ket{0}-\frac{1}{\sqrt{2}}\ket{1} 
& \mbox{if $b=0$, $x=1$} \cr
\frac{1}{\sqrt{2}}\ket{0}+\frac{1}{\sqrt{2}}\ket{2} 
& \mbox{if $b=1$, $x=0$} \cr
\frac{1}{\sqrt{2}}\ket{0}-\frac{1}{\sqrt{2}}\ket{2} 
& \mbox{if $b=1$, $x=1$}
\end{array}\right.\]

\begin{enumerate}
\item
Alice picks a uniformly random $b\in\{0, 1\}$ and $x\in\{0, 1\}$
and sends $\ket{\phi_{b, x}}$ to Bob.
\item
Bob picks a uniformly random $b'\in\{0, 1\}$, sends $b'$ to Alice.
\item
Alice sends $b$ and $x$ to Bob, he checks if 
the state that he received from Alice in the $1^{\rm st}$ step
is $\ket{\phi_{b, x}}$ (by measuring it in with respect to
in a basis consisting of $\ket{\phi_{b, x}}$ and 
two vectors orthogonal to it)\footnote{For example, if $b=x=0$,
then Bob could measure in the basis 
$\ket{\phi_{00}}=\frac{1}{\sqrt{2}}\ket{0}+\frac{1}{\sqrt{2}}\ket{1}$,
$\frac{1}{\sqrt{2}}\ket{0}-\frac{1}{\sqrt{2}}\ket{1}$, $\ket{2}$.}.
If the outcome of the measurement is not $\ket{\phi_{b, x}}$,
he has caught Alice cheating and he stops the protocol.
\item
Otherwise, the result of the coin flip is $b\oplus b'$. 
\end{enumerate}

\begin{Theorem}
\label{Protocol}
The bias of this protocol is $0.25$.
\end{Theorem}

\noindent{\bf Proof:}
We bound the probability of dishonest Alice (or dishonest Bob)
achieving $b\oplus b'=0$. The maximum probability of achieving
$b\oplus b'=1$ is the same because the protocol is symmetric.

{\em Case 1: Alice is honest, Bob cheats.}
If $b=0$, Alice sends a mixed state that is equal to 
$\frac{1}{\sqrt{2}}(\ket{0}+\ket{1})$ with probability 1/2
and $\frac{1}{\sqrt{2}}(\ket{0}-\ket{1})$ with probability 1/2.
If $b=1$, she sends a mixed state that is equal to
$\frac{1}{\sqrt{2}}(\ket{0}+\ket{2})$ with probability 1/2
and $\frac{1}{\sqrt{2}}(\ket{0}-\ket{2})$ with probability 1/2.
The density matrices of these two mixed states are
\[ \rho_0=\left( 
\begin{array}{ccc} 
\frac{1}{2} & 0 & 0 \\
0 & \frac{1}{2} & 0 \\
0 & 0 & 0 \end{array} 
\right) 
\rho_1=\left( 
\begin{array}{ccc} 
\frac{1}{2} & 0 & 0 \\
0 & 0 & 0 \\
0 & 0 & \frac{1}{2} 
\end{array}
\right) 
\]
and $\|\rho_0-\rho_1\|_t=1$.
By Theorem 3 of \cite{ATVY}, 
the probability that Bob achieves $b=b'$ is
at most $\frac{1}{2}+\frac{\|\rho_0-\rho_1\|_t}{4}=\frac{3}{4}$.

{\em Case 2: Bob honest, Alice cheats.}

Let $\rho$ be the density matrix of the state sent by Alice in the
$1^{\rm st}$ step. 
The first step of the proof is to ``symmetrize'' 
Alice's strategy so that it becomes easier to bound her
success probability. 

\begin{Lemma}
\label{Symmetrize}
There is a strategy for dishonest Alice where the state sent by Alice
in the $1^{\rm st}$ round has the density matrix of the form
\begin{equation}
\label{eq1} \rho'=
\left(
\begin{array}{ccc}
1-\delta_1-\delta_2 & 0 & 0 \\
0 & \delta_1 & 0 \\
0 & 0 & \delta_2
\end{array} \right) 
\end{equation}
for some $\delta_1$ and $\delta_2$ and Alice achieves $b=b'$ 
with the same probability.
\end{Lemma}

\noindent{\bf Proof:}
Let $U_0=I$, 
\[ U_1=
\left(
\begin{array}{ccc}
1 & 0 & 0 \\
0 & -1 & 0 \\
0 & 0 & 1
\end{array} \right)~~~ 
U_2=\left(
\begin{array}{ccc}
1 & 0 & 0 \\
0 & 1 & 0 \\
0 & 0 & -1
\end{array} \right)
\]\[
U_3=\left(
\begin{array}{ccc}
1 & 0 & 0 \\
0 & -1 & 0 \\
0 & 0 & -1
\end{array} \right). \]
Assume that Alice, before sending the state $\ket{\psi}$ to Bob in the 
$1^{\rm st}$ round, applies $U_i$ to it and, then,
in the $3^{\rm rd}$ round, replaces each description of 
$\ket{\phi_{b, x}}$ by a description of
$U_i\ket{\phi_{b, x}}$. 
Then, Alice achieves the outcomes 0 and 1 and gets caught with the same
probabilities as before because
\begin{enumerate}
\item[(a)]
For all $i\in\{0, 1, 2, 3\}$, $b\in\{0, 1\}$, $x\in\{0, 1\}$,
$U_i\ket{\phi_{b, x}}$ is either $\ket{\phi_{b, 0}}$ or 
$\ket{\phi_{b, 1}}$, and
\item[(b)]
For any $\ket{\psi}$, the inner product between $U_i\ket{\psi}$ and
$U_i\ket{\phi_{b, x}}$ is the same as the inner product between
$\ket{\psi}$ and $\ket{\phi_{b, x}}$.	 
\end{enumerate}

Probabilities of obtaining 0, 1 and getting caught also stay the same if
Alice picks a uniformly random $i\in\{0, 1, 2, 3\}$ and then applies $U_i$
to both the state sent in the $1^{\rm st}$ round and the description
sent in the $3^{\rm rd}$ round.
In this case, the density matrix of the state sent by Alice in the
$1^{\rm st}$ round is $\rho'=\frac{1}{4} (
U_0\rho U_0^{\dagger}+
U_1\rho U_1^{\dagger}+ U_2\rho U_2^{\dagger}+
U_3\rho U_3^{\dagger})$.
For every $j, k\in\{1, 2, 3\}$, $j\neq k$,
$(U_i \rho U^{\dagger}_i)_{jk}$ is equal to 
$\rho_{jk}$ for two $i\in\{0, 1, 2, 3\}$ and
to $-\rho_{jk}$ for the two other $i$.
Therefore, $\rho'_{jk}=0$ for all $j\neq k$,
i.e. $\rho'$ is of the form (\ref{eq1}).
\qed

\begin{Lemma}
\label{Alice}
For ``symmetrized'' Alice's strategy,
the probability that Alice convinces Bob that $b=0$ is at most
$F(\rho', \rho_0)$.
\end{Lemma}

%\noindent{\bf Note:}
%The proof uses properties of Alice's strategy
%which hold because of ``symmetrization''. 
%B%ecause of thathe lemma might not be true if the honest protocol 
%used a different set of states $\ket{\phi_{ax}}$.

\noindent{\bf Proof:}
Let
\begin{equation}
\label{new-eq} 
\ket{\psi}=\sum_{i} a_i \ket{i}\ket{\psi_i} 
\end{equation}
be the purification of $\rho'$ chosen by Alice 
if she want to convince Bob that $b=0$.
For every $\ket{\psi_i}$, Alice sends to Bob a description of a state
$\ket{\psi'_i}$ which is one of $\ket{\phi_{b, x}}$, $b\in\{0, 1\}$,
$x\in\{0, 1\}$.

Alice is trying to convince Bob that $b=0$.
Therefore, we can assume that she always sends to Bob
a description of $\ket{\phi_{0, 0}}$ or $\ket{\phi_{0, 1}}$.
(Replacing a description of $\ket{\phi_{1, x}}$ by a description of
$\ket{\phi_{0, x}}$ can only increase the probability of Bob 
accepting $b=0$, although it may simultaneously increase
the probability of Alice caught cheating.)

We pair up each state $\ket{\psi_i}$ with the state 
$\ket{\psi_j}=U_1\ket{\psi_i}$
and each state $U_2\ket{\psi_i}$ with $U_3\ket{\psi_i}=U_2 U_1\ket{\psi_i}$.
Our ``symmetrization'' guarantees that 
\begin{itemize}
\item 
if $\ket{\psi_i}$ and $\ket{\psi_j}$ are the
two states in one pair, then $a_i=a_j$,
\item
if one of states in a pair has $\ket{\psi'_i}=\ket{\phi_{0,0}}$,
the other has $\ket{\psi'_j}=U_1\ket{\phi_{0,0}}=\ket{\phi_{0,1}}$, and
conversely,
\item
$\lbra \psi_i | \psi'_i\rket=\lbra \psi_j|\psi'_j\rket$ 
(because performing $U_1$ maps $\ket{\psi_i}$ and $\ket{\psi_j}$
to $\ket{\psi'_i}$ and $\ket{\psi'_j}$, respectively).
\end{itemize}
Therefore, we can write the equation (\ref{new-eq}) as
\begin{equation}
\label{new-eq1} 
\ket{\psi}=\sum_{i} a_i \left( \frac{1}{\sqrt{2}}\ket{i,0}\ket{\psi_{i, 0}}
+\frac{1}{\sqrt{2}}\ket{i,1}\ket{\psi_{i, 1}} \right)  
\end{equation}
with $\ket{\psi'_{i,0}}=\ket{\phi_{0,0}}$ and 
$\ket{\psi'_{i,1}}=\ket{\phi_{0,1}}$.

The probability that Bob accepts $\ket{\psi_{i,x}}$ as $\ket{\psi'_{i,x}}$
is $|\lbra \psi_{i,x} | \psi'_{i,x}\rket|^2$.
The total probability of Bob accepting is
\begin{equation}
\label{new-eq2} 
\sum_i \frac{1}{2} |a_{i}|^2 (|\lbra \psi_{i,0} | \psi'_{i,0} \rket|^2 +
|\lbra \psi_{i,1} | \psi'_{i,1} \rket|^2 ).
\end{equation}
Notice that, because of ``symmetrization'', 
$\lbra \psi_{i, 0}|\psi'_{i, 0}\rket = \lbra\psi_{i, 1}|\psi'_{i, 1}\rket$.
Therefore, if we define 
$\ket{\varphi_i}=\frac{1}{\sqrt{2}}\ket{i,0}\ket{\psi_{i, 0}}
+\frac{1}{\sqrt{2}}\ket{i,1}\ket{\psi_{i, 1}}$ and
$\ket{\varphi'_i}=\frac{1}{\sqrt{2}}\ket{i,0}\ket{\psi'_{i, 0}}
+\frac{1}{\sqrt{2}}\ket{i,1}\ket{\psi'_{i, 1}}$,
we have $\lbra \varphi_i | \varphi'_i\rket=\lbra \psi_{i,0}|\psi'_{i, 0}\rket=
\lbra\psi_{i,1}|\psi'_{i,1}\rket$.
This means that (\ref{new-eq2}) is equal to
\[ \sum_i |a_{i}|^2 |\lbra \varphi_i | \varphi'_i\rket|^2 .\]
Let $\rho_i$ be a mixed state which is $\ket{\psi_{i,0}}$
with probability 1/2 and $\ket{\psi_{i,1}}$
with probability 1/2.
Then, $\rho'=\sum_i |a_i|^2 \rho_i$.
Since $\ket{\varphi_i}$ and $\ket{\varphi'_i}$ are purifications 
of $\rho_i$ and $\rho_0$, we have 
$|\lbra\varphi_i | \varphi'_i\rket|^2 \leq F(\rho_i, \rho_0)$
and
\[ \sum_i |a_{i}|^2 |\lbra \varphi_i | \varphi'_i\rket|^2 \leq 
\sum_i |a_i|^2 F(\rho_i, \rho) .\]
By concavity of fidelity \cite{NC}, 
\[ \sum_i |a_i|^2 F(\rho_i, \rho_0) \leq F(\sum_i |a_i|^2 \rho_i, \rho_0)
= F(\rho, \rho_0) .\]
\qed

\begin{Lemma}
The probability that Alice achieves $b\oplus b'=0$ (or, equivalently,
$b\oplus b'=1$) is at most $\frac{1}{2}(F(\rho', \rho_0)+F(\rho', \rho_1))$.
\end{Lemma}

\noindent{\bf Proof:}
With probability 1/2, Bob's bit is $b'=0$. 
Then, to achieve $b\oplus b'=0$, Alice needs to convince him that $b=0$.
By Lemma \ref{Alice}, she succeeds with probability at most
$F(\rho', \rho_0)$.

With probability 1/2, Bob's bit is $b'=1$.
Then, Alice needs to convince Bob that $b=1$ and
she can do that with probability $F(\rho', \rho_1)$.
The overall probability that Alice succeeds is 
$\frac{1}{2}(F(\rho', \rho_0)+F(\rho', \rho_1))$.
\qed

By Lemma \ref{ULemma}, 
\[ F(\rho',
\rho_0)=[Tr(\sqrt{\sqrt{\rho'}\rho_0\sqrt{\rho'}})]^2 =
\left(\frac{1}{\sqrt{2}}\sqrt{1-\delta_1-\delta_2}+
\frac{1}{\sqrt{2}}\sqrt{\delta_1}\right)^2. \]
Similarly, $F(\rho', \rho_1)=
(\frac{1}{\sqrt{2}}\sqrt{1-\delta_1-\delta_2}+
\frac{1}{\sqrt{2}}\sqrt{\delta_2})^2$.
Therefore,
\[ \frac{1}{2}(F(\rho', \rho_0)+F(\rho', \rho_1)) \]
\[ =(\frac{1}{\sqrt{2}}\sqrt{1-\delta_1-\delta_2}+
\frac{1}{\sqrt{2}}\sqrt{\delta_1})^2+
(\frac{1}{\sqrt{2}}\sqrt{1-\delta_1-\delta_2}+
\frac{1}{\sqrt{2}}\sqrt{\delta_2})^2 \]
\begin{equation}
\label{eq2} 
= \frac{1}{2}\left(
(1-\delta_1-\delta_2) + \frac{\delta_1}{2} + \frac{\delta_2}{2}
+ \sqrt{1-\delta_1-\delta_2}(\sqrt{\delta_1}+\sqrt{\delta_2})\right) .
\end{equation}
Let $\delta=\frac{\delta_1+\delta_2}{2}$.
The convexity of the square root implies that 
$\sqrt{\delta_1}+\sqrt{\delta_2}\leq 2\sqrt{\delta}$ and (\ref{eq2}) is at 
most
\[ \frac{1}{2} \left(1-\delta+2\sqrt{\delta(1-2\delta)}\right) .\]

Taking the derivative of this expression shows that it is maximized
by $\delta=\frac{1}{6}$.
Then, it is equal to $\frac{1}{2}
(1-\frac{1}{6}+\frac{4}{6})=\frac{3}{4}$.
\qed

%The best previously known bound was $\frac{\sqrt{8}-1}{2}=0.41...$\cite{ATVY},
%using a protocol with a similar structure but different set of states
%$\ket{\phi_{b, x}}$.
%Our proof technique can be used to improve the bound for their protocol
%to $\frac{1}{\sqrt{2}}=0.35...$.

\section{Lower bound for 3 rounds}
\label{3roundlb}

%We have two lower bounds for 3-round protocols.
We show a lower bound for a class of
3 round protocols which includes the protocol of section \ref{section2}
and the protocol of \cite{ATVY}. %(but not the protocol of \cite{Salvail}).
This class is defined by fixing the structure of the protocol
and varying the choice of states $\ket{\phi_{b, x}}$.

Let $X_0$ and $X_1$ be two sets and $\pi_0$ and $\pi_1$ be
probability distributions over $X_0$ and $X_1$, respectively.
Assume that, for every $b\in\{0, 1\}$ and $x\in X_b$ we have
a state $\ket{\phi_{b, x}}$.

\begin{enumerate}
\item
Alice picks a uniformly random $b\in\{0, 1\}$.
Then, she picks $x\in X_b$ according to the 
distribution $\pi_b$ and sends $\ket{\phi_{b, x}}$ to Bob.
\item
Bob picks a random $b'\in\{0, 1\}$, sends $b'$ to Alice.
\item
Alice sends $b$ and $x$ to Bob.
Bob checks if the state that he received %from Alice 
in the $1^{\rm st}$ step
is $\ket{\phi_{b, x}}$.
\item
The result of the coin flip is $b\oplus b'$. 
\end{enumerate}

%The next theorem shows that the protocol of Theorem \ref{Protocol}
%is optimal for this class. 

\begin{Theorem}
\label{Restricted}
Any protocol of this type has a bias at least 
0.25.
\end{Theorem}

\noindent{\bf Proof:}
Let $\rho_0$ and $\rho_1$ be the density matrices sent by an honest Alice
if $b=0$ and $b=1$, respectively.
(These density matrices are mixtures of $\ket{\phi_{b, x}}$ over
$x\in X_i$.)

\begin{Lemma}
\label{BLemma}
Bob can achieve 0 with probability 
$\frac{1}{2}+\frac{\|\rho_0-\rho_1\|_t}{4}$.
\end{Lemma}

\noindent{\bf Proof:}
By Lemma \ref{AKNLemma}, there is a measurement $\M$ that, 
applied to $\rho_0$ and $\rho_1$, produces two 
probability distributions with the variational distance between them
equal to $\|\rho_0-\rho_1\|_t$ and it can be chosen so that there are 
just two outcomes: 0 and 1.

Let $p_0$ and $1-p_0$ be the probabilities of outcomes 0 and 1 when
the measurement $\M$ is applied to $\rho_0$.
For the variational distance to be $\|\rho_0-\rho_1\|_t$,
the probabilities of outcomes 0 and 1 when the measurement $\M$ 
is applied to $\rho_1$ have to be $p_0-\frac{\|\rho_0-\rho_1\|_t}{2}$
and $1-p_0+\frac{\|\rho_0-\rho_1\|_t}{2}$.

Bob applies the measurement $\M$ to the state that he receives from Alice
and sends $b=0$ if the measurement gives 0 and $b=1$ if the measurement 
gives 1. Since an honest Alice chooses $a=0$ with probability 1/2 and
$a=1$ with probability 1/2, Bob achieves $a=b$ (and $a\oplus b=0$)
with probability 
\[ \frac{1}{2} p_0 + \frac{1}{2} 
\left(1-p_0+\frac{\|\rho_0-\rho_1\|_t}{2}\right)=
\frac{1}{2}+\frac{\|\rho_0-\rho_1\|_t}{4} .\]
\qed

\begin{Lemma}
\label{ALemma}
Alice can achieve 0 with probability  \[ \frac{1}{2}+
\frac{\sqrt{F(\rho_0, \rho_1)}}{2}. \]
\end{Lemma}

\noindent{\bf Proof:}
First, we consider an honest Alice which does the protocol
on a quantum level. That means that she flips a classical coin to
determine $a\in\{0, 1\}$ and then prepares the superposition 
\[ \ket{\psi_a}=\sum_{i\in X_a} \sqrt{\pi_a(i)} \ket{i} \ket{\phi_{a, i}} \]
and sends the second part of the superposition to Bob.
After receiving $b$ from Bob, she measures $i$ and sends $a$ and $i$ to Bob.

%A dishonest Alice can cheat by preparing a state $\ket{\varphi_0}$
%between $\ket{\psi_0}$ and $\ket{\psi_1}$ and then, after receiving $b$,

The pure states $\ket{\psi_0}$ and $\ket{\psi_1}$ are purifications of
the density matrices $\rho_0$ and $\rho_1$.
By Lemma \ref{JLemma}, there is a unitary transformation $U$
on the Alice's part of $\psi_1$ such that 
$|\lbra \psi_0 | U(\psi_1) \rket^2 =F(\rho_0, \rho_1)$.

Let $\alpha$ be such that $F(\rho_0, \rho_1)=\cos^2 \alpha$.
Then, $\lbra \psi_0 | U(\psi_1) \rket=\cos\alpha$.
This means that
\[
\left\{
\begin{array}{l}
\ket{\psi_0} = \cos\frac{\alpha}{2} \ket{\varphi_0} + 
 \sin\frac{\alpha}{2} \ket{\varphi_1} \\
U \ket{\psi_1} = \cos\frac{\alpha}{2} \ket{\varphi_0} - 
 \sin\frac{\alpha}{2} \ket{\varphi_1}
\end{array} 
\right.
\]
for some states $\ket{\varphi_0}$, $\ket{\varphi_1}$.

A dishonest Alice prepares $\ket{\varphi_0}$ 
and sends the $2^{\rm nd}$ part to Bob.
If she receives $b'=0$ from Bob, she acts as an honest quantum Alice 
who has prepared $\ket{\psi_0}$ and sent the $2^{\rm nd}$ part to Bob.
(That is, she measures her part $\ket{i}$ and sends 0 and $i$ to Bob.)
Bob accepts $b=0$ with probability at least\footnote{For 
a formal proof of this, define $\H$ to be the space
of all bipartite states $\ket{\psi}$ such that Bob accepts with
probability 1, if Alice acts in this way.
Then, $\ket{\psi_0}\in \H$ and, since the angle between 
$\ket{\varphi_0}$ and $\ket{\psi_0}$ is $\frac{\alpha}{2}$,
the squared projection of $\ket{\varphi_0}$ on $\H$ is at least 
$\cos^2\frac{\alpha}{2}$, implying that Bob accepts with probability
at least $\cos^2 \frac{\alpha}{2}$ if Alice starts with $\ket{\varphi_0}$.}
$|\lbra \psi_0 | \varphi_0\rket|^2=\cos^2 \frac{\alpha}{2}$.

If she receives $b'=1$, Alice performs $U^{-1}$ on her part
of $\ket{\varphi_0}$ and continues as an honest quantum Alice who
has prepared $\ket{\psi_1}$ (measures $\ket{i}$ and sends to 1 and $i$ to
Bob). 
Bob accepts $b=1$ with probability 
\[ |\lbra U^{-1}(\varphi_0) | \psi_1 \rket|^2 =
|\lbra \varphi_0 | U(\psi_1) \rket|^2 = \cos^2 \frac{\alpha}{2}. \]

In both cases, the probability of Bob accepting $b\oplus b'=0$ is
$\cos^2 \frac{\alpha}{2}$. 
Therefore, the overall probability of $b\oplus b'=0$ is 
$\cos^2\frac{\alpha}{2}$ as well and we have
\[ \cos^2\frac{\alpha}{2}=\frac{1+\cos\alpha}{2} = 
\frac{1+\sqrt{F(\rho_0, \rho_1)}}{2} .\]
\qed

If $F(\rho_0, \rho_1)\geq \frac{1}{4}$, then, by Lemma \ref{ALemma},
Alice can achieve a bias of 
$\frac{\sqrt{F(\rho_0, \rho_1)}}{2}\geq \frac{1}{4}$.

If $F(\rho_0, \rho_1)\leq \frac{1}{4}$, then, by Lemma \ref{BLemma}, 
Bob can achieve a
bias of $\frac{1}{4}\|\rho_0-\rho_1\|_t$ and, by Lemma \ref{FLemma},
\[ \frac{1}{4} \|\rho_0, \rho_1\|_t \geq 
  \frac{1}{2} (1-\sqrt{F(\rho_0, \rho_1)}) \geq
  \frac{1}{4} .\]
\qed

\section{Two standard forms for quantum protocols}
\label{SF}

We use the following ``standard form'' for quantum protocols.
%
%state two standard forms for quantum coin flipping protocols.
%Any protocol can be transformed into a protocol in any one
%of these two forms without increasing the bias $\epsilon$.

\begin{Theorem}
\cite{Mayers}
\label{Standard1}
If there is a protocol for quantum coin flipping with a bias at most
$\epsilon$, there is also a protocol with bias at most $\epsilon$ in which no
party makes measurements until all communication is complete.
\end{Theorem}

The main idea of the proof of Theorem \ref{Standard1} is that
all measurements can be delayed till the end of protocol.
For more details, see Mayers \cite{Mayers}.
This result has been very useful for proving
the impossibility of quantum bit commitment \cite{LC,Mayers}
and other cryptographic primitives.  
%We will also use it in the next section.

A different ``standard form'' has been pointed out
to us by Kitaev \cite{Kitaev0}.
% of an opposite type.
%Instead of replacing measurements and classical communication
%by unitary transformations and quantum communication everywhere,
%we want to decrease the quantum communication as much as possible.
%Obviously, to have a quantum protocol, we need at least one
%round of quantum communication. It turns out that this
%is enough.

\begin{Theorem}
\label{Standard2}
\cite{Kitaev0}
If there is a protocol for quantum coin flipping with a bias $\epsilon$,
there is another protocol with bias $\epsilon$ in which only the first
message from Alice to Bob is quantum and all the other messages
are classical.
\end{Theorem}

The proof idea is that Alice transmits a lot of EPR pairs in
the first round and, after that, Alice and Bob can 
replace all quantum communication by classical communication
using teleportation.
%The result is new. It was pointed out to us by Kitaev \cite{Kitaev0}.

We do not use this result in our paper but we
decided to mention it because it might be useful
for other purposes.
%It does not have any applications yet but it might be 
%useful in the future, for the following reason.
The protocols of the form of Theorem \ref{Standard2}
have a fairly simple structure.
In the first step Alice creates an entangled
state with Bob and then they both do operations on their qubits
and communicate classical information. 
Because of this simple structure, they might be easier 
to analyze than general protocols. 
We note that this structure of a protocol somewhat resembles
the well-known LOCC (local operations and classical communication)
paradigm in the study of entanglement \cite{Nielsen,NC}.

\section{The lower bound on the number of rounds}
\label{section1}

\begin{Theorem}
\label{Rounds}
Let $\epsilon<1/4$.
Any protocol for quantum coin flipping that achieves a bias
$\epsilon$ must use $\Omega(\log \log \frac{1}{\epsilon})$ rounds.
\end{Theorem}

%\subsection{Proof overview}

%\noindent
%{\bf Proof sketch:}
Assume we have a protocol for quantum coin flipping with $k$ rounds
and a bias $\epsilon$.
By Theorem \ref{Standard1}, we can assume
that this protocol does not make any measurements till the
end of communication.

The protocol starts with a fixed starting state $\ket{\psi^0}$.
Then (if both players are honest), 
Alice applies a unitary transformation $U_1$, sends some
qubits to Bob, he applies $U_2$, sends some qubits to Alice and so on.
After $U_k$, both Alice and Bob perform measurements on their parts.
Since there is no measurements till the communication is finished,
the joint state of Alice and Bob after $i$ steps is a pure state
$\ket{\psi^i}$.

At the end of protocol, Alice and Bob measure the final state to
determine the outcome of the coin flip.
If both Alice and Bob follow the protocol, the two measurements give
the same result and this result is 0 with probability 1/2 and 1
with probability 1/2. 
%For the purpose of 
%our analysis, we assume that this final
%measurement only measures the 0/1 result bit and does not disturb 
%other qubits.
%We also assume that all intermediate measurements are delayed
%till the end of the protocol. This is possible because of the
%``principle of safe storage'' of \cite{BV}.

%Then, the joint state of Alice and Bob after $i$ steps is a pure state
%$\ket{\psi^i}$.
% joint state of Alice and Bob after $i$ rounds
%and $\rho^i_{A}$, $\rho^i_{B}$ be the density matrices of 
%Alice's and Bob's part, respectively.

For our analysis, we decompose each of intermediate states $\ket{\psi^i}$ as
$\ket{\psi^i_0}+\ket{\psi^i_1}$,
where $\ket{\psi^i_0}$ is the state which leads to the outcome 0
if the rest of protocol is applied and $\ket{\psi^i_1}$ is
the state which leads to the outcome 1
if the rest of protocol is applied.
This is done as follows.

First, we decompose the final state $\ket{\psi^k}$.
Let $\ket{\psi^k_0}$ and $\ket{\psi^k_1}$ 
be the (unnormalized) states after the final measurement 
if the measurement gives 0 (1). 
Then, $\ket{\psi^k_0}\perp\ket{\psi^k_1}$ and
$\ket{\psi^k}=\ket{\psi^k_0}+\ket{\psi^k_1}$.
Also, $\|\psi^k_0\|^2 = \|\psi^k_1\|^2 = \frac{1}{2}$ (since a protocol
must give 0 with probability 1/2 and 1 with probability 1/2).

Next, we define $\ket{\psi^i_0}=(U_{i+1} U_{i+2} \ldots U_k)^{-1}
\ket{\psi^k_0}$ and $\ket{\psi^i_1}=(U_{i+1} U_{i+2} \ldots U_k)^{-1}
\ket{\psi^k_1}$.
Then, $\ket{\psi^k} = \ket{\psi^k_0}+\ket{\psi^k_1}$ %and
%the linearity of $U_{i+1}\ldots U_k$ 
implies $\ket{\psi^i} = \ket{\psi^i_0}+\ket{\psi^i_1}$. 

Let $\rho^i_{A, j}$ ($\rho^i_{B, j}$) be the density matrix
of Alice's (Bob's) part of the (normalized) 
bipartite state $\sqrt{2}\ket{\psi^i_j}$.
Let $F^i_A$ ($F^i_B$) be the fidelity between $\rho^i_{A, 0}$
and $\rho^i_{A, 1}$ ($\rho^i_{B, 0}$ and $\rho^i_{B, 1}$).

Our proof is based on analyzing how $F^i_A$ and $F^i_B$
change during the protocol. We show that they must be large at
the beginning, 0 at the end and, if they decrease too
fast, this creates an opportunity for cheating.
This implies the lower bound on the number of rounds.

We start with the simplest part of the proof:
$F^i_A$ and $F^i_B$ must be small at the end ($i=k$).

\begin{Lemma}
\label{EndLemma}
$F^k_A=F^k_B=0$.
\end{Lemma}

\noindent {\bf Proof:}
At the end, both Alice and Bob know the outcome of the protocol
with certainty. That means that there is a measurement of
Alice's qubits that perfectly distinguishes $\ket{\psi^k_0}$
and $\ket{\psi^k_1}$ (i.e., this measurement gives 0 with probability
1 on $\ket{\psi^k_0}$ and 1 with probability $\ket{\psi^k_1}$).  
By Lemma \ref{AKNLemma}, $\|\rho^k_{A, 0}-\rho^k_{A, 1}\|_t=2$.
By Lemma \ref{FLemma}, $F(\rho^k_{A, 0}, \rho^k_{A, 1})$ must be 0.

Similarly, $F^k_B=0$.
\qed

Second, we show that, if $F^0_A$ or $F^0_B$ is too small,
one of sides can cheat.

\begin{Lemma}
\label{StartLemma}
Alice can achieve one of outcomes 0 and 1 with probability 
at least $1-\sqrt{F^0_A}$.
\end{Lemma}

\noindent {\bf Proof:}
Since there is no communication before the start of the protocol, 
the starting superposition
$\ket{\psi^0}$ is a tensor product $\ket{\psi_A}\otimes\ket{\psi_B}$,
with Alice having $\ket{\psi_A}$ and Bob having $\ket{\psi_B}$.

Consider the best measurement $\M$ (for Alice)
that distinguishes $\rho^0_{A, 0}$
and $\rho^0_{A, 1}$ (Lemma \ref{AKNLemma}).
Then, $\ket{\psi^0_0}=\ket{\psi_{00}}+\ket{\psi_{01}}$,
where $\ket{\psi_{00}}$ is the remaining state if 
the measurement $\M$ on $\ket{\psi^0_0}$ gives the outcome 0
and $\ket{\psi_{01}}$ is the remaining state if 
$\M$ gives the outcome 1.
Let $\ket{\psi^0_1}=\ket{\psi_{10}}+\ket{\psi_{11}}$,
with $\ket{\psi_{10}}$ and $\ket{\psi_{11}}$ defined similarly.

If Alice applies $\M$ to 
$\ket{\psi^0}=\ket{\psi^0_0}+\ket{\psi^0_1}$,
she either gets the outcome 0 and the remaining state 
$\ket{\psi'_0}=\ket{\psi_{00}}+\ket{\psi_{10}}$ or 1 and the remaining state
$\ket{\psi'_1}=\ket{\psi_{01}}+\ket{\psi_{11}}$.
$\ket{\psi^0}$ is a product state and the measurement $\M$ is
applied to Alice's side only.
Therefore, $\ket{\psi'_0}$ and
$\ket{\psi'_1}$ (the remaining states when $\M$ gives 0 and 1)
are product states as well.

Since $\ket{\psi^0}=\ket{\psi'_0}+\ket{\psi'_1}$,
either $\|\psi'_0\|^2\geq \frac{1}{2}$
or $\|\psi'_1\|^2\geq \frac{1}{2}$.
For simplicity, we assume that $\|\psi'_0\|^2\geq \frac{1}{2}$
and Alice is trying to achieve the outcome 0.
(The outcome 1 can be achieved similarly 
with a slightly smaller probability.)

Let $\ket{\psi'_A}\otimes \ket{\psi_B}$ be the normalized state 
$\frac{\ket{\psi'_0}}
{\|\psi'_{0} \|}$.
To bias the coin towards 0, Alice just runs the 
honest protocol with her starting state being
$\ket{\psi'_A}$ instead of $\ket{\psi_A}$.

Let $\|\psi_{01}\|^2+\|\psi_{10}\|^2\leq \epsilon$.
We show that this implies that $\ket{\psi'_A}\otimes\ket{\psi_B}$
is close to the normalized state $\sqrt{2}\ket{\psi^0_0}$ 
(which gives the outcome 0 with probability 1).
We have 
\[ \frac{\ket{\psi'_{0}}}{\|\psi'_{0}\|} =
\frac{\ket{\psi_{00}}+\ket{\psi_{10}}}{\|\psi'_0\|} =
\frac{\ket{\psi_{00}}+\ket{\psi_{01}}}{\|\psi'_0\|} +
\frac{\ket{\psi_{10}}-\ket{\psi_{01}}}{\|\psi'_0\|} .\]
$\ket{\psi_{00}}+\ket{\psi_{01}}=\ket{\psi^0_0}$ leads
to the outcome 0 with certainty.
Therefore, the probability of a different outcome
(1 or Alice caught cheating) is at most
\[ \frac{\|\psi_{10}-\psi_{01}\|^2}{\|\psi'_0\|^2} \leq 
\frac{\|\psi_{10}\|^2+\|\psi_{01}\|^2}{\|\psi'_0\|^2}
\leq \frac{\epsilon}{1/2}=2\epsilon. \]
Therefore, the described strategy for dishonest Alice
gives 0 with probability at least $1-2 \epsilon$.

Next, we bound $\|\psi_{01}\|^2+\|\psi_{10}\|^2$.

Let $1-p_0$ and $p_0$ be the probabilities of outcomes 0 and 1 when 
measuring $\sqrt{2}\ket{\psi^i_0}$. 
Let $p_1$ and $1-p_1$ be the probabilities
of 0 and 1 when measuring $\sqrt{2}\ket{\psi^i_1}$.  
Then, the variational distance between these two probability
distributions is $2(1-p_0-p_1)$.
Since we are using the best measurement for
distinguishing $\rho^0_{A, 0}$ and $\rho^0_{A, 1}$,
$2(1-p_0-p_1)$ is equal to $\|\rho^0_{A, 0}-\rho^0_{A, 1}\|_t$.
By Lemma \ref{FLemma}, this implies 
\[ 1-\sqrt{F^0_A} = 
1-\sqrt{F(\rho^0_{A, 0}, \rho^0_{A, 1})} \leq 
\|\rho^0_{A, 0}-\rho^0_{A, 1}\|_t =
(1-p_0-p_1) \]
and this is equivalent to $p_0+p_1\leq \sqrt{F^0_A}$.

Notice that $p_0=2\|\psi_{01}\|^2$ because $\rho^0_{A, 0}$ is the
density matrix of Alice's side of $\sqrt{2}\ket{\psi^0_{A, 0}}$ and
$\ket{\psi_{01}}$ is the remaining state if the measurement of
$\ket{\psi^0_{A, 0}}$ gives 1.
Similarly, $p_1=2\|\psi_{10}\|^2$.
Therefore, we have 
$\|\psi_{01}\|^2+\|\psi_{10}\|^2 \leq \frac{1}{2} \sqrt{F^0_A}$
and Alice can bias the coin to 0 with
probability at least $1-\sqrt{F^0_A}$.
\qed

Hence, if the bias of a protocol is $\epsilon$, then, 
by Definition \ref{SecDef}, we must
have $1-\sqrt{F^0_A}\leq \frac{1}{2}+\epsilon$.
This implies $\sqrt{F^0_A}\geq \frac{1}{2}-\epsilon$ and
$F^0_A\geq (\frac{1}{2}-\epsilon)^2$.
Since $\epsilon<1/4$, we must have 
$F^0_A\geq \frac{1}{16}$. 

Third, we show that, if after any round,
one of $F^i_A$ and $F^i_B$ is much larger than the other,
this also creates a possibility for cheating. 

\begin{Lemma}
\label{Main}
Let $i\in\{1, \ldots, k-1\}$.
Then, there is a strategy for dishonest Alice 
which achieves the result 0 with probability at least
\begin{equation}
\label{eq4} 
\left( \frac{1}{\sqrt{2}} - \sqrt[4]{F^i_A} \right)^2+
 \left( \frac{\sqrt{F^i_B}}{\sqrt{2}} - \sqrt[4]{F^i_A} 
\right)^2 .
\end{equation}
\end{Lemma}

%Therefore, whenever $\frac{\delta_B}{2}>2\sqrt[4]{\delta_A}$,
%Alice can successfully cheat. Similarly, if
%$\frac{\delta_A}{2}>2\sqrt[4]{\delta_B}$, then Bob
%can cheat.

\noindent {\bf Proof:}
For brevity, we denote $F^i_A$ and $F^i_B$ by
$F_A$ and $F_B$ (omitting the index $i$ which is the same
throughout the proof).

We first prove the $F_A=0$ case. 
This case was previously considered by Mayers et.al.\cite{MSC}.
They showed that, if $F_A=0$ and $F_B>0$, then Alice can
successfully cheat. 
Below, we show how to formalize their argument so that
it shows the probability that Alice can achieve.

\noindent
{\em $F_A=0$ case.}
Then, (\ref{eq4}) is just $\frac{1}{2}+\frac{F_B}{2}$.

By Lemma \ref{FLemma}, $F(\rho^i_{A, 0}, \rho^i_{A, 1})=F_A=0$ implies
$\|\rho^i_{A, 0}-\rho^i_{A, 1}\|_t=2$. By Lemma \ref{AKNLemma},
there is a measurement for Alice that perfectly distinguishes  
$\rho^i_{A, 0}$ and $\rho^i_{A, 1}$.
%Since this measurement
%Alice can perform this measurement without disturbing the rest of
%the state, i.e. so that the joint state of Alice and Bob after the
With probability 1/2, the outcome of the measurement is 0 and
the joint state of Alice and Bob after the measurement
is $\ket{\psi^i_0}$.
With probability 1/2, the outcome is 1 and
the joint state of Alice and Bob
becomes $\ket{\psi^i_1}$.
%
%measurement is $\ket{\psi^i_0}$ with probability 1/2 and $\ket{\psi^i_1}$ 
%with probability 1/2.
In the first case, she just continues as in the honest protocol.
This gives the answer 0 with probability 1/2.

If she gets $\ket{\psi^i_1}$, by Lemma \ref{JLemma}, 
there is a unitary transformation $U$
that can be performed by Alice such that 
\begin{equation}
\label{eq0} 
| \lbra \psi^i_0 | U(\psi^i_1) \rket |^2 = 
F(\rho^i_{B, 0}, \rho^i_{B, 1}) \|\psi^i_0\|^2 =
\frac{F(\rho^i_{B, 0}, \rho^i_{B, 1})}{2} = 
\frac{F_B}{2} . 
\end{equation}
Alice performs $U$ and then continues
as in the honest protocol. 
This gives the answer 0 with probability at least $F_B/2$.

Together, the probability of answer 0 is at least 
$\frac{1}{2} (1+F_B)$.

\medskip

\noindent
{\em $F_A\geq 0$ case.}
By Lemma \ref{FLemma}, 
there is a measurement $\M$ for Alice that, applied to $\rho^i_{A, 0}$
and $\rho^i_{A, 1}$, produces two probability 
distributions with the variational distance 
between them at least $2(1-\sqrt{F_A})$.
Without the loss of generality, we can assume that this is
a measurement with two outcomes 0 and 1 and the probability of
0 is higher for $\rho^i_{A, 0}$ and the probability of 1 
is higher for $\rho^i_{A, 1}$.

The strategy for cheating Alice is the same as in the $F_A=0$ case.
She applies the measurement $\M$ and, then, if she gets 0, continues
as in the honest protocol.
If she gets 1, she applies the transformation $U$ and then
continues as in the honest protocol.

Next, we show that this strategy achieves the result 0 with the probability
given by the formula (\ref{eq4}).

%Let $1-p_0$ and $p_0$ be the probabilities of outcomes 0 and 1 when 
%measuring $\rho^i_{A, 0}$.
%Let $p_1$ and $1-p_1$ be the probabilities
%of 0 and 1 when measuring $\rho^i_{A, 1}$.  
%Then, the variational distance is $2(1-p_0-p_1)$ and
%$2(1-p_0-p_1)\geq 2(1-\sqrt{\delta_A})$ is equivalent to
%$p_0+p_1\leq \sqrt{F_A}$.

%Then, on $\ket{\psi^i_{A, 0}}$, this measurement must give 0 with
%probability at least $1-\sqrt{\delta_A}$ and, on $\ket{\psi^i_{A, 1}}$, it
%must give 1 with probability at least $1-\sqrt{\delta_B}$.

Let $\ket{\psi'_0}$ and $\ket{\psi'_1}$ denote the (unnormalized) 
remaining states when the outcome of the measurement $\M$
is 0 and 1, respectively.

Also, let $\ket{\psi_{ab}}$ (for $a, b\in\{0, 1\}$)
denote the (unnormalized) remaining
states when $\ket{\psi^i_{a}}$ is measured and 
the outcome of the measurement is $b$. 
Since $\ket{\psi^i}=\ket{\psi^i_0}+\ket{\psi^i_1}$, we have
$\ket{\psi'_0}=\ket{\psi_{00}}+\ket{\psi_{10}}$ and
$\ket{\psi'_1}=\ket{\psi_{01}}+\ket{\psi_{11}}$.

On the other hand, $\ket{\psi^i_0}=\ket{\psi_{00}}+\ket{\psi_{01}}$ and
$\ket{\psi^i_1}=\ket{\psi_{10}}+\ket{\psi_{11}}$.
%Notice that $\| \psi'_{01}\|^2=\frac{p_0}{2}$ 
%(because $p_0$ is the probability of obtaining 1 when
%measuring the normalized state $\sqrt{2}\ket{\psi^i_0}$).
%Similarly, $\| \psi'_{10}\|^2=\frac{p_1}{2}$. 
Therefore,
\begin{equation}
\label{eq11}
\| \psi'_0 - \psi^i_0 \| = 
 \| \psi_{10}-\psi_{01} \| \leq \| \psi_{10} \| + 
 \| \psi_{01}\| 
 \leq 
\sqrt{2 ( \|\psi_{10}\|^2 + \|\psi_{01}\|^2 )}  .
\end{equation}
Similarly to the proof of Lemma \ref{StartLemma},
$\|\psi_{10}\|^2+\|\psi_{01}\|^2 \leq \frac{1}{2}\sqrt{F_A}$.
Therefore, (\ref{eq11}) is at most $\sqrt[4]{F_A}$.
%By Cauchy-Schwartz inequality, this is at most
%\[ \sqrt{\frac{\|\psi'_{10}\|^2+\|\psi'_{01}\|^2}{2}}=
%\sqrt{\frac{p_0+p_1}{4}} =\frac{\sqrt{p_0+p_1}}{2} \leq
%\frac{\sqrt[4]{\delta_A}}{2}. \]
%$\|\ket{\psi'_{10}}\|$ and $\|\ket{\psi'_{01}}\|$ are at most
%$\sqrt[4]{\delta_A}$ each (because the probabilities of
%obtaining 1 from measuring $\ket{\psi^i_0}$ or 0 from
%measuring $\ket{\psi^i_1}$ are $\|\ket{\psi'_{10}}\|$ and 
%$\|\ket{\psi'_{01}}\|$ and they must be at most $\sqrt{\delta_A}$
%each). 
We also have
$\| \psi'_1-\psi^i_1\|\leq \sqrt[4]{F_A}$
with the same proof.

Let $\H^i_0$ be the set of bipartite states such that applying the rest of
the protocol ($U_k U_{k-1}\ldots U_{i+1}$) and the final measurement
at the end of the protocol gives the outcome 0 with probability 1.
Then, $\ket{\psi^i_0}\in \H^i_0$.
Also, the norm of the projection of $U(\ket{\psi^i_1})$ on 
$\H^i_0$ is at least $\sqrt{F_B/2}$ (by (\ref{eq0})).

Consider the norms of the projections of $\ket{\psi'_0}$ and $\ket{\psi'_1}$
on $\H^i_0$.
They differ from the norms of $\ket{\psi^i_0}$ and $\ket{\psi^i_1}$
by at most $\|\psi'_0-\psi^i_0\|\leq \sqrt[4]{F_A}$ 
and $\|\psi'_1-\psi^i_1\|\leq \sqrt[4]{F_A}$.
Therefore, the projection of $\ket{\psi'_0}$ on $\H^i_0$ is
of norm at least $\frac{1}{\sqrt{2}}-\sqrt[4]{F_A}$ and
the projection of $U\ket{\psi'_1}$ is of norm at least
$\frac{\sqrt{F_B}}{\sqrt{2}}-\sqrt[4]{F_A}$.
This means that the probability of outcome 0
is at least
\[ \left( \frac{1}{\sqrt{2}} - \sqrt[4]{F_A} \right)^2+
 \left( \frac{\sqrt{F_B}}{\sqrt{2}}-
\sqrt[4]{F_A} \right)^2 .\]
%Therefore,
%\[ \| \ket{\psi'_0} - \frac{1}{\sqrt{2}} \ket{\psi^i_0} \| \leq
%\sqrt{2} \sqrt[4]{\delta_A} ,\]
%\[ \lbra \psi'_0 | \psi^i_0\rket = \lbra\psi^i_0 | \psi^i_0\rket+
%\lbra\psi'_0-\psi^i_0 | \psi^i_0 \rket \geq
%\frac{1}{2} - \| \psi'_0-\psi^i_0\| \cdot \|\psi^i_0\| \geq
%\frac{1}{2} -\sqrt[4]{\delta_A} \]
\qed

For the purposes of this paper, a weaker form of lemma \ref{Main}
is sufficient.

\begin{Corollary}
Let $i\in\{1, \ldots, k-1\}$.
Then, there is a strategy for dishonest Alice 
which achieves the result 0 with probability at least
$\frac{1}{2}+\frac{F_B}{2}-2\sqrt{2}\sqrt{F_A}$.
\end{Corollary}

\noindent {\bf Proof:}
We have
\[ \left( \frac{1}{\sqrt{2}} - \sqrt[4]{F_A} \right)^2+
 \left( \frac{\sqrt{F_B}}{\sqrt{2}}-
 \sqrt[4]{F_A} \right)^2 \] \[ =
\left( \frac{1}{2} - \sqrt{2} \sqrt[4]{F_A}+
 \sqrt{F_A} \right) +
\left( \frac{F_B}{2} - 
\sqrt{2}\sqrt{F_B}\sqrt[4]{F_A} +
 \sqrt{F_A} \right) \]
\[ \geq \left( \frac{1}{2} - 
\sqrt{2} \sqrt[4]{F_A} \right)+
  \left( \frac{F_B}{2} - \sqrt{2} \sqrt[4]{F_A} \right) \] \[ =
\frac{1}{2} + \frac{F_B}{2} -2\sqrt{2}\sqrt[4]{F_A} .\]
\qed

\begin{Corollary}
\label{MainCor}
Assume that the bias of a protocol is at most $\epsilon$.
Then, after every round, $F_B\leq 2\epsilon+6\sqrt[4]{F_A}$
and $F_A\leq 2\epsilon+6\sqrt[4]{F_B}$.
\end{Corollary}

\noindent {\bf Proof:}
By Lemma \ref{Main}, Alice can achieve
$Pr[0]=\frac{1}{2}+\frac{F_B}{2}-2\sqrt{2}\sqrt[4]{F_A}$.
Because the bias of the protocol is at most $\epsilon$,
we must have $\frac{F_B}{2}-2\sqrt{2}\sqrt[4]{F_A}\leq \epsilon$
and $F_B\leq 2\epsilon+4\sqrt{2}\sqrt[4]{F_A}\leq
2\epsilon + 6\sqrt[4]{F_A}$.

$F_A \leq 2\epsilon + 6\sqrt[4]{F_B}$ follows similarly.
\qed

Next, we use Corollary \ref{MainCor} to show that 
the fidelities $F^i_A$ and $F^i_B$ cannot decrease too fast.

\begin{Lemma}
\label{Induction}
Assume that a $k$-round protocol has the bias at most $\epsilon$.
Then, for any $i<k$, $F^i_A\leq 14 \epsilon^{1/4^{k-i-1}}$
and $F^i_B\leq 14 \epsilon^{1/4^{k-i-1}}$.
\end{Lemma}

\noindent {\bf Proof:}
By induction on $k-i$.

{\em Base case.} $i=k-1$.

First, remember that $F^k_A=F^k_B=0$.
Let $X\in\{A, B\}$ be the person who sends the message in the $k^{\rm th}$
round and $Y$ be the person who receives the message.
Sending away a part of the state can only increase the fidelity.
Therefore, $F^{k-1}_X\leq F^k_X=0$, i.e. $F^{k-1}_X=0$.

By Corollary \ref{MainCor}, $F^{k-1}_Y \leq 2\epsilon+6\sqrt[4]{F^{k-1}_X}
= 2\epsilon < 14\epsilon$. 
  
{\em Inductive case.}

We assume that the lemma is true for $i$ and show that it is
also true for $i-1$.
Similarly to the previous case, let $X$ be the person who sends the
message in the $i^{\rm th}$ round and $Y$ be the other person.
Then, 
\[ F^{i-1}_X\leq F^i_X \leq 14 \epsilon^{1/4^{k-i-1}} .\]
By Corollary \ref{MainCor},
\[ F^{i-1}_Y\leq 2\epsilon+6\sqrt[4]{14 \epsilon^{1/4^{k-i-1}}} \leq
(2+6\sqrt[4]{14}) \epsilon^{1/4^{k-i}} < 14 \epsilon^{1/4^{k-i}} .\]
\qed

In particular, Lemma \ref{Induction} implies that 
$F^0_A\leq 14 \epsilon^{1/4^{k-1}}$.
We also have $F^0_A\geq \frac{1}{16}$ (Lemma \ref{StartLemma}
and the first paragraph after its proof). 
Therefore,
$14 \epsilon^{1/4^{k-1}}\geq \frac{1}{16}$.
Taking log of both sides twice gives
$k=\Omega(\log \log \frac{1}{\epsilon})$.

\section{Conclusion}

We have constructed a protocol for quantum coin flipping with
bias 0.25 and shown that it is optimal for a restricted class 
of protocols. We also gave a general lower bound on the number of
rounds needed to achieve a bias $\epsilon$.
A stronger lower bound has been later shown by Kitaev \cite{Kitaev}
but our bound also applies to weak coin flipping. 
The table below summarizes the known results for both strong
and weak coin flipping.

\medskip

\begin{tabular}{|l|l|l|}
\hline
 & Best protocol & Best lower bound \\
\hline
Strong & 0.25 (this paper) & $\frac{1}{\sqrt{2}}-\frac{1}{2}=0.21...$ 
\cite{Kitaev} \\
\hline
Weak & $\frac{1}{\sqrt{2}}-\frac{1}{2}=0.21...$ \cite{RS} & $\epsilon>0$, 
$\Omega(\log\log \frac{1}{\epsilon})$ rounds required \\
& &(this paper)\\
\hline
\end{tabular}

\medskip

The bounds for strong coin flipping are quite close but 
weak coin flipping is still wide open. 

Another interesting question about coin flipping
protocols is ``cheat-sensitivity'' studied by \cite{ATVY,Kent,RS}.
A protocol is for coin flipping or other cryptographic tasks
is cheat-sensitive if a dishonest party cannot
increase the probability of one outcome without
being detected with some probability. 
Many quantum protocols display some cheat-sensitivity
but it remains to be seen what degree of cheat-sensitivity
can be achieved.

\section{Acknowledgments}

Thanks to Dorit Aharonov, Daniel Gottesman, Alexei Kitaev, Boaz Leslau, 
Hoi-Kwong Lo, Rashindra Manniesing, Moni Naor, 
Ashwin Nayak, Louis Salvail, Yaoyun Shi, Rob Spekkens, Amnon Ta-Shma, 
Umesh Vazirani and Xinlan Zhou for useful comments, discussions
and information about related work.

\end{document}